\documentclass[8pt]{article}
\usepackage{amsgen,amsmath,amstext,amsbsy,amsopn,amssymb}
\usepackage{graphicx}
\usepackage[utf8]{inputenc}
\usepackage{amsthm}
\RequirePackage{bm}
\usepackage{mathrsfs}
\usepackage{mathtools}
\usepackage{bbm}
\usepackage{subcaption}
\usepackage{float}
\usepackage{comment}

\usepackage{hyperref}
\usepackage[square,sort,comma,numbers]{natbib} 

\usepackage{rotating}
\usepackage{titling}
\usepackage{algorithmic}
\usepackage{xcolor}

\usepackage{authblk}

\usepackage{graphicx}
\usepackage{url}

\newcommand{\tb}{\mathbf{t}}

\newcommand{\Vb}{\mathbf{V}}

\newcommand{\Zb}{\mathbf{Z}}

\newcommand{\Prob}{\mathbb{P}}
\newcommand{\Expec}{\mathbb{E}}
\newcommand{\Var}{\text{Var}}

\definecolor{mygrey}{gray}{0.45}

\newcommand{\ch}[1]{\textcolor[rgb]{0,0,0}{{#1}}}
\newcommand{\chsb}[1]{\textcolor[rgb]{0,0,0}{{#1}}}
\newcommand{\jf}[1]{\textcolor[rgb]{0,0,0}{{#1}}}
\newcommand{\fixit}[1]{\textcolor[rgb]{0,0,0}{{#1}}}
\newcommand{\lastpass}[1]{\textcolor[rgb]{0,0,0}{{#1}}}

\usepackage{rotating}

\begin{document}
	
	\begin{title}
		{\Large\bf Intertemporal Community Detection in Human Mobility Networks}
	\end{title}
	
	\author[1]{Mark He \footnote{MH was funded by government support under contract FA9550-11-C-0028 and awarded by the Department of Defense, Air Force Office of Scientific Research, National Defense Science and Engineering Graduate (NDSEG) Fellowship, 32 CFR 168a ; Corresponding Author: Corresponding email at markhe@live.unc.edu}}
	\author[1]{Joseph Glasser}
	\author[1]{Shankar Bhamidi PhD \footnote{SB was supported in part by NSF grants DMS-1613072, DMS-1606839 and ARO grant W911NF-17-1-0010.}}
	\author[2]{Nikhil Kaza PhD}
	
	\renewcommand\Affilfont{\small}
	
	\affil[1]{Statistics \& Operations Research, University of North Carolina at Chapel Hill, Chapel Hill, NC, 27599, USA}
	\affil[2]{City \& Regional Planning, University of North Carolina at Chapel Hill, Chapel Hill, NC, 27599, USA}
	\maketitle

\begin{abstract}
 We introduce a community detection method that finds clusters in network time-series by introducing an algorithm that finds significantly interconnected nodes across time. 
 These connections are either  increasing, decreasing, or constant over time. 
 Significance of nodal connectivity within a set  is judged using the Weighted Configuration Null Model at each time-point, then a novel significance-testing scheme is used to  assess  connectivity at all time points and the direction of its time-trend. We apply this method to bikeshare networks in New York City and Chicago and taxicab pickups and dropoffs in New York to find and illustrate patterns in human mobility  {in} urban zones. Results show stark geographical patterns in clusters that are growing and declining in relative  usage across time and potentially elucidate latent economic or demographic trends.
\end{abstract}
 


\section{Introduction} \label{sec:intro}

\ch{Much research has been done in recent years in the analysis of real world networks. One particular area of active interest is \textit{community detection}. Broadly, community detection is an unsupervised exploratory data analysis method that extracts subsets of vertices in a network that are more densely connected within the subset than between the subsets in a given network. }

\ch{A majority of the research on community detection \jf{in} networks has dealt with static networks \citep{girvan_community_2002}. However, many real-world networks exhibit dynamic properties, such as human mobility networks in urban systems. These networks include commuting patterns over time \citep{patuelli2010evolution}, location based social networks \citep{assemSpatioTemporalClusteringApproach2016}, taxicab travel patterns \citep{liuRevealingTravelPatterns2015a} and cell phone call records \citep{Reades2009}.  }  Understanding the structures of these networks \jf{reveals} underlying trends in  human mobility  and provides important information \jf{for the management of urban infrastructure.}

There are many human mobility patterns \jf{that can be represented} as networks with high temporal resolution because of the presence of origin and destination locations and time stamps associated with the trips. \ch{ For example, bikeshare systems are rich and remarkably comprehensive in tracking mobility patterns within a city.} By 2019, over 2,000 cities have created bikeshare systems around the world. In 2018, according to the National Association of City Transportation Officials, 36.5 million trips were completed in over 100 cities in the United \lastpass{States using} these systems. Many of these systems have stations where users can rent the bikes and deposit them at another station at the end of the trip. These stations allow the system operator to track the precise origins and destinations of individual trips by time-of-day and day-of- week. Another mode of travel in cities is by automobile, which trips can be modeled as networks. In particular, taxicabs in  cities are regulated and therefore location and time data of these cab pickup and dropoff locations are often reported to the regulators. The increased usage of often less-regulated ridesharing services (Uber, Lyft etc.) have reduced taxi trips in the last few years. Much research has been done on network analyses \cite{austwick_structure_2013, cazabet_grav} \cite{Zhan_taxi_graph2016,tong_taxi_2017}, but most do not fully take into account the dependencies induced by the network structures and temporal trends. Many of these studies have also focused mostly on demand estimation \cite{10.1371/journal.pone.0137922,FAGHIHIMANI201553}.

In this work, we develop a method to identify clusters of significantly connected nodes in a time-series of weighted networks. Identification of such clusters \jf{allows} us to understand the nature of geographical, economic and cultural relationships, when these networks are rooted in cities. \ch{ Identifying trajectories of connectivity in clusters across time may reveal structural changes within the mobility patterns in the city.}   \ch{ We develop an intertemporal community detection method to analyze the structure of long-term trends in time-series of networks to understand \textit{global} and \textit{local} trends. In particular, we attempt to determine whether such trends are uniformly distributed across the networks, or whether certain communities exhibit countervailing trends in interconnectivity \jf{when compared with} others.  We aim to identify and partition the nodes that potentially have driven this global trend, as well as the communities that exhibit locally specific trends.}

\ch{ The objective of the community detection method in this study is to find groups of nodes that are consistently connected across time and \jf{exhibit increasing, decreasing, or stable trends} in connectivity.} We use the assumptions of the weighted configuration model as posited in \cite{palowitch_continuous,He2019} to \textit{scale} and \textit{whiten} the time-series of node-set connectivities. In doing so, we remove much of the overall graph effects that represent network-wide signals at a given time-point, such as weather and other city-wide phenomena. Though normalization removes much of the seasonality, some autoregressivity still \jf{persists}: we ignore these effects but future work should account for such behavior. Analysis of time-varying weighted graphs  \jf{allows} us to gain more insight \jf{into} the nature of the city as a complex \jf{accumulation} of micro-level spatial activity patterns.  While this method of intertemporal community detection is developed for data structured like mobility systems, it can be adapted for any type of time-series network data with registered nodes (such as inter county commuting patterns, internet traffic, etc.)

\section{Data and Network Construction}

We apply intertemporal community detection to data from two bikeshare systems and a taxicab trips.  \ch{Bikeshare trip data for Divvy (Chicago) and Citibike (New York) are publicly available on their respective websites \cite{divvy_data},  \cite{citibike}}.  The two bikeshare systems provide \jf{contrasting cases}.  Divvy ridership increased steadily between 2014-2016 from 2.7 to 3.6 million, but overall ridership declined slightly from  3.8 million trips in 2017 to  $\sim$3.6 million in 2018 \cite{greenfield_2018}.  The Citibike system, on the other hand, has consistently increased in usage from 14 million in 2016 to 16 million in 2017 and 18 million in 2018  \cite{citibike}. 

The publicly available datasets include trip start and stop times for each trip between stations. In our analyses, we focus on the time period between July 2016 and June 2018. We omit all stations that were newly introduced or removed within this period. 547 nodes (7.4 million trips) in Chicago and 583 nodes (8.4 million trips) in New York remain in the dataset used for this study. One common problem in bikeshare systems is the issue of \textit{supply-demand mismatch} in ridership. A station in a high-activity area of a large city is often empty or full at certain times of the day \cite{Gast:2015:PFB:2806416.2806569, XIE_DemandAnalysis,pendem2019,Freund2018MinimizingMF,FaghihImani2014AnalysingBS, 10.1371/journal.pone.0137922}. A full or empty station prevents an otherwise possible trip. Load rebalancing is a well-studied problem for bikeshare systems in order to solve the inefficiencies associated with queuing between bikes in stations with finite numbers of slots for bikes at each station.  Real-time data on station status \jf{rebalancing exist} for New York and Chicago  \cite{divvy_gfbs}. However, historical station inventory data  is only available for New York City \cite{openbus} and not Chicago. Thus, for the New York bikeshare system, we find communities with and without demand adjustment (see section \ref{sec:demand_correction} for details on the method). 

The taxicab data for New York is from the Taxicab and Limosuine Commission \cite{tlc2019}. We use data from January 2017 to 2019 because trips from the ridehailing apps (such as Uber, Jio) are only included since 2017 in the data. 263 zones cover all the five boroughs of New York, and the dataset includes over 453 million trips between these zones. 

From these datasets, we construct the observed time-series of networks as $\{G_t \}_{1 \le t \le T}$. In all these datasets, we aggregate the trips between a pair of nodes for each week. The weekly aggregation smooths the diurnal variations and keeps the  time-series long enough for time-domain. Thus, each time $t$ corresponds to a week, where $T$ is the total number of time periods. The indicator $A_{uv,t}$ represents the presence of any trips at time $t$ between $u$ and $v$. We use the number of trips \jf{between two} nodes per week $t$ as the edge weight $W_{uv,t} $. In network $G_t$  the degree of node $u$  is defined as $\deg_{u,t} = \sum_{v: v \ne u} A_{uv,t} $ and strengths are defined as $S_{u,t} = \sum_{v: v \ne u} W_{uv,t} $ at each time-unit $t$ across total time $T$ \cite{He2019}. We define the index set $[n]$ as the set of all nodes $u$, which represent stations.

\section{Detecting Intertemporal Communities}

In this section, we describe a  method based on iterative testing of node-set connectivities to extract statistically significant communities across time \cite{He2019, palowitch_continuous}. 
We use a similar approach but account for and classify the types of time dependency. 
We posit that trends across time are generally \textit{increasing, decreasing} or \textit{stable} and account for these types of time dependence. To this end, we adjust connectivities to time-decay and find trends using equivalence testing \cite{Schuirmann1987, dixon_trend_2008}.

We introduce a method to find clusters that are  significantly connected across time and exhibit differing trends in connectivity by building on the iterative testing framework of Palowitch et al.(\cite{palowitch_continuous}). In that work, the weights on the edges incident on each node $u$ are modeled as
\begin{eqnarray}\label{eq: null_model} 
\widehat{W}_{uv}=
\xi_{uv}  \bigg(\frac{s_u s_v}{s_T} \bigg) \big/  \bigg( \frac{d_u d_v}{d_T} \bigg)  
\end{eqnarray}

where each $d_u$  represents the  degree of node $u$, where each $s_{u}$ represents the strength from node $u$, and $s_T, d_T$  represents the global sum of strengths and degrees.  

\textit{Communities} are sets of vertices that have edges that are significantly interconnected \textit{within} the set but not connected outside the set. Prior work \cite{palowitch_continuous, He2019} use the above null model to identify communities within a single graph.   Random variables $\xi_{uv}$ with variance $\kappa$ are constructed so as to satisfy  the weighted configuration model  \cite{He2019,palowitch_continuous}.  When $W_{uv,t}$ are structured in time-series, the variances $\kappa_t$ of random variable $\xi_{uv,t}$ summarize the overall variability of each network across time. We utilize these time-varying characteristics of the networks  to find intertemporal communities.

\subsection{Intertemporal Configuration Null Model}  \label{sec:intertemporal_null}

We extend the method used in prior work \cite{palowitch_continuous,He2019} by developing an intertemporal null model wherein a baseline model is extracted from a time-series of registered networks and iteratively  subjected to hypothesis tests for trends and local deviance. 
We detect significant communities across the time-series of networks if the trend  \textbf{and}  variation components are significantly different  from those of the baseline model.	These communities signal subsections of the network that are either strongly interconnected at either the beginning or end of the time-period,  or consistently connected throughout the entire time period. 

  We define the intertemporal null model for a given node set $B$  (as in \cite{palowitch_continuous, He2019}) to determine if it  is  significantly interconnected across   \textit{all} time-points according to the hypothesized trend. 
	We search for communities that are
\begin{itemize}
	\item \textit{increasing} if its nodes are significantly connected at time $t=1$, but not necessarily significantly connected as $t$ is  late.
	\item \textit{decreasing} if its nodes are significantly connected at time $t=T$, but not necessarily significantly connected when $t$ is early.
	\item \jf{\textit{stable}} (or neutral) if its nodes are significantly connected across all time points.
\end{itemize}
Within $B$, we posit that a time-series of relative connectivity may be decomposed into \textit{trend} and \textit{variation} components. Trend denotes the presence of a constant  time-trend in the relative connectivity amongst nodes in set $B$.  Variation denotes the aspects of the node-set connectivity that do not vary systematically across time.

\subsubsection{Null Model for Node-Set Connectivity}

The estimate  for each edge weight $W_{uv,t}$ at time $t$ is a simple extension of \eqref{eq: null_model}, which is the null model for a single graph. 

\begin{align}\label{eq: inter_null_model} 
\widehat{W}_{uv,t}=
\begin{cases}
\xi_{uv,t}  (\frac{s_{u,t} s_{v,t}}{s_{T,t}}) \big/  (\frac{d_{u,t} d_{v,t}}{d_{T,t}})  & \text{if }  u \ne v \\
0  & \text{if }  u = v
\end{cases} 
\end{align}

Each $\widehat{W}_{uv,t}$  is an weighted edge on a random time-varying graph $\mathcal{G}_t$, where each $u$ has fixed degrees $d_{u,t}$ and strengths $s_{u,t}$.
 Each graph $G_t$ at time $t$ has  total degrees $d_{T,t} = \sum_v d_{v,t}$  and total strengths $s_{T,t} = \sum_v s_{v,t}$. The analogous node-set connectivity $S(u,B, {G}_t) $  (as \cite{palowitch_continuous}),   is
\begin{equation}\label{eq: S_uB}
S(u,B, {G}_t)  = \sum_{v \ne u, v \in B} W_{uv, t}  
\end{equation}

with means and variances 
\begin{align} \label{eq:ccme_meanvars}
\Expec[S(u,B,\mathcal{G}_t)]  &=    \sum_{v \in B} \frac{s_{u,t} s_{v,t}}{ s_{T,t}}  ; \\
\Var( S(u,B ,\mathcal{G}_t))  &=   \sum_{v \in B}     \frac {  (\frac{s_{u,t} s_{v,t}}{ s_{T,t} } )^2} {  \frac{d_{u,t} d_{v,t}}{ d_{T,t} } } \left(  \kappa_t -   \frac{d_{u,t} d_{v,t}}{d_{T,t}} + 1 \right).
\end{align} 
detailed derivations of these values can be found in the text of \cite{palowitch_continuous}.

To search for intertemporal clusters, we investigate the significance of connectivities of $B_t$ to $v_t$  across time $t$. 
The normalized connectivity score $Z_t(v,B)$ represents the sums of weights of a given node set $B$ in graphs $ G_t $:

\begin{equation}\label{eq:NormalizeduB}
Z_t (v,B) =    \frac{ S(v,B, {G}_t) - \Expec[ S (v,B, \mathcal {G}_t) ]  }{\Var( S(v,B, \mathcal{G}_t)) }   , \quad   { t = 1,...,T} 
\end{equation} 


\subsection{Identifying Nodes that are Significantly Bordering  Across Time } \label{sec:significant_nodes}

We use iterative testing to identify nodes that are significantly connected to their neighbors through time.  
Methods developed in previous literature \citep{palowitch_continuous, He2019} have applied this method to a fixed graph $G$. 
We use the same method of deriving significance of the probability that  $v$is significantly connected to $u$  in set $B$ as in Palowitch et al. (\cite{palowitch_continuous}).

The proposed algorithm relies on an iterative procedure starting at iteration step $k=1$, then repeated until the results do not change.
The objective is to find sets $B$  such that for each $v \in B$, $v$ is significantly connected to $u$ across \textit{all} time points $T$. 
At a given step $k>1$,  for fixed time $t$, for a set of nodes $B_{k,t}$ and a bordering node $u$, the score of node-set connectivity is determined by \eqref{eq: S_uB}:
\begin{equation}
S(u,B_{k,t}, {G}_t)  = \sum_{v \ne u, v \in B_{k,t}} W_{uv, t}.  
\end{equation}
After the normalizing calculation \eqref{eq:NormalizeduB}  is performed,  a p-value for each $v  \in B_{k,t}$ is then determined:
\begin{equation}\label{eq: pvalue}
p(u,B_{k,t}, {G}_t ) =   \Prob (   S(u,B_{k,t}, {G}_t) >   S(u,B_{k,t}, \mathcal{G}_t ) ).
\end{equation} 

For each time point $t$, the p-value  $p(u,B_{k,t}, {G}_t )$  is then corrected for false-discovery rate correction as in \cite{wilson_essc}.  The non-significant nodes are rejected and  the a set of significant nodes is retained.	We describe additional steps to find significant nodes in the following sections \ref{sec: ordinary_BH} - \ref{sec: Bonferroni_interval} to account for time-decay in significant bordering nodes and describe the testing of trends in \ref{sec:trend_test}. 

\subsubsection{Time-Decay Adjusted False Discovery Rate Correction} \label{sec: ordinary_BH}
To identify significantly interconnected nodes for a given time $t$, a slightly augmented version of the  Benjamini-Hochberg \cite{bh_reject} procedure used in \cite{palowitch_continuous,He2019} is used. The only difference is that the FDR-adjusted p-value $p^*_u$ is  multiplied by  decay term $a_t$, contingent on if the communities are hypothesized to be  increasing, decreasing, or stable in connectivity over time.
For a fixed time $t$, iteration step $k$, and set $B_{k,t}$, we find all the nodes that are significantly connected to $B_{k,t}$  across all time $t=1,...,T$ after calculating the p-value as in \eqref{eq: pvalue}.
The output set at iteration $K$ and time $t$ is written as $M_k(B_k)$, described in more detail in later sections in equation \eqref{eq:M_def}.  
 

We define $a_t$ is an exponential decay term to adjust for the shifting time-window of significance. It is  defined as:
\begin{eqnarray}
a_t  :=
\begin{cases}
\left(  1 -  \exp \left( - \frac{ t-1 }{T}  \right)  \right) a_0^+    & \text{if trend is increasing} \\
\left(   \exp \left( - \frac{ t-1 }{T}\right)  - a^-_0  \right) \bigg/ (1-a^-_0)      & \text{ if trend is decreasing }\\ 
1 & \text{if trend is neutral.}
\end{cases}
\end{eqnarray}
The terms $ a_0^+ $ and   $a_0^-$ are defined such that $a_t$ is  0 at time 1 and 1 at time $T$ if the trend is increasing, and 1 at time $1$ and 0 at time $T$ if the trend is decreasing:
\begin{align*}
a_0^+    :=  1 -  \exp \left(-\frac{T-1}{T}  \right) ; \quad \quad a_0^-    :=    \exp \left(-\frac{T-1}{T}  \right) .
\end{align*}

If the trend is posited to be increasing, then the algorithm allows more permissive selection of `significantly' bordering nodes when time $t$ is early, but is more penalizing when $t$ approaches $T$.  
When $t$ is 1, then  $a_t$ is equal to zero. In this case, all $p^*_u$ are zero and will automatically be counted as significant if $u$ borders $B_{k,t}$. When $t$ is $T$, then $	a_t $ is 1, so the FDR correction is identical to BH.
{The  threshold for the maximum allowable p-value increases  as $t$ decreases so that negligible  connections (when  $t$ is early) that become stronger (when $t$ is late) are deemed significant.} 

Conversely, when the trend is posited to be decreasing, the same kind of correction is made in reverse because  the   multiplier is subtracted by one. Because the multiplier to the adjusted p-value is always less than 1, the procedure is always less conservative than the Benjamini-Hochberg method and lets nodes that otherwise would not be significant at a given time-period be deemed as``significant" based on their potential to be significant given their trajectory. If the trend is posited to be neutral, then we use the ordinary BH  rejection procedure.

\subsubsection{Bonferroni Interval for Bordering Frequencies}
\label{sec: Bonferroni_interval}

\newcommand{\Btk}{ B_{k,t}}

The previous section \ref{sec: ordinary_BH} details significance  testing  for the  collections of nodes $\Btk$ at each time period. 
To determine whether the collections of nodes are significantly connected to $u$ at \textit{all} times, we apply a second testing step using Bonferroni Correction. 
This correction is  applied  to the frequencies of nodes whose p-values have been deemed significant by the BH correction (in the previous section). 
The product of Bonferroni confidence intervals is used to define  {the} significance of  {the} neighboring frequency at iteration step $k$, for each $B_{k,t} $  across all time $t=1,...,T$.
For  a set $\Btk$, we define $m_t(\Btk)$ as the set of nodes that are found to be `significantly bordering' described by Section \ref{sec: ordinary_BH}:
$$m_t(\Btk) = \# \{ \text{$u$: $u$ is significantly bordering $\Btk$ at time  } t   \}.$$
A large value of $m_t(B_{k,t})$ for all $t$ signifies a large collection of nodes that significantly border $B_{k,t}$ and  results in a false discovery interval that is close to $T$, and hence $v$ must border $B_{k,t}$ for nearly all time $T$ for it to be significant.
Conversely, if  $m_t(B_{k,t})$ is small, then the required frequency for $v$ to be significant is not as high. During this step, we assume  away dependency between  $B_{k,t}$.

We define $FDI_{\alpha,k}$ to be the threshold for false discovery interval of all significantly adjacent nodes to node $B_{k,t}$
$$ FDI_{\alpha,k} =   \prod_{t =1,..., T}\left(  1 - \frac{\alpha} {m_t(\Btk)} \right) \cdot T $$
where 	$ 1 - {\alpha}/ {m_t(\Btk)} $ is the Bonferroni confidence level at each time point. 
The product of  these intervals cross all time multiplied by the total time $T$ gives the threshold of significantly bordering nodes across all time. 
 
\begin{figure}[htbp]
	\centering
	\includegraphics[width=.49\linewidth   ,   trim={2cm 21cm 13cm 2cm},clip]{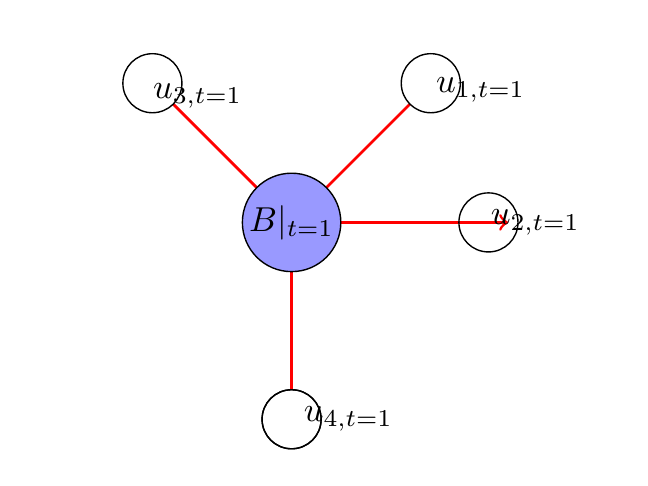}
	\includegraphics[width=.49\linewidth, trim={2cm 21cm 13cm 2cm},clip]{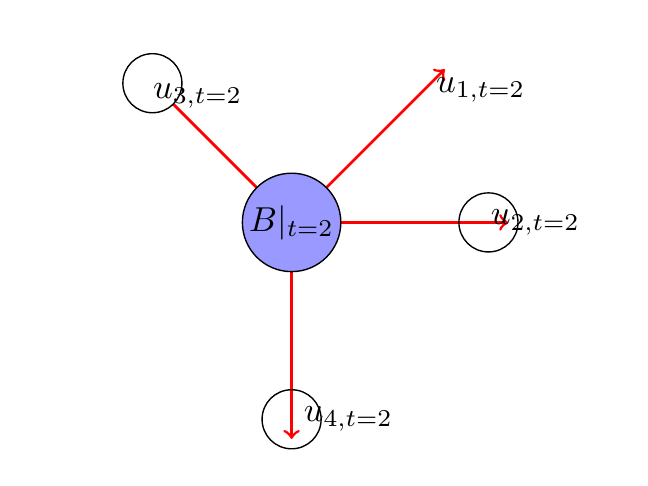}
		\caption{\small Example of set $B$ at times $t=1,2$. $u_1$ is significantly connected when $t=1$, but not when $t=2$. So for arbitrary iteration step $k$, let $B_k = B$, then ${m_t(\Btk)}$ is ${m_1(B_{1,k})}= B_k \bigcup \{  u_1, u_2, u_3, u_4 \}$ at $t=1$ , but ${m_2(B_{2,k})}= B_k \bigcup \{  u_2, u_3,u_4 \}$ at $t=2$.   }
\end{figure}


Now we define the $B_k^0$ as the combined list of all the nodes in any $B_{k,t}$:
$B^0_k = \bigcup_{t  = 1,...,T} B_{k,t}. $  For each $v \in B^0_k$, we define the bordering frequency $N_v (B_k)$  as the counts of $v$ which are significantly bordering $\Btk$ across all time $t$.
A significant  $N_v(B_k)$  suggests that  $v$ is more frequently bordered across time than other nodes. 
Each $v$ significantly borders all $\Btk$   if 
\begin{equation}  \label{eq:significantly_neighboring}
FDI_{\alpha, k}< N_v(\Btk)
\end{equation}
 that is, if $\Btk$ borders $v$ enough times across $t$ for it to be significant overall in the time-period $1,...,T $ \citep{dunn1959}. Finally,  we take the union of all nodes $v$ that satisfy the ``significantly neighboring" criteria \eqref{eq:significantly_neighboring} and denote the set  $M_k(B_k)$ 
\begin{equation} \label{eq:M_def}
M_k(B_k) = \bigcup_{v \in B^0_k } \{ v:  	FDI_{\alpha, k}  < N_v(B_k)\}.
\end{equation}


The resulting set $M_k(B_k) $ {represents} the nodes that are significantly connected across time, given the appropriate time-window adjustments. 
We then check if the trends are actually as hypothesized.

\subsection{\textbf{Significance Testing for Trends}}  \label{sec:trend_test}

We  define the sum of   $Z_t (v,B) $ in \eqref{eq:NormalizeduB}   as $\Zb (B)$ to gauge the significance of the time-trend of a cluster.
\begin{align*}   
\Zb (B) &=  \biggl\{  \sum_{v \in B} Z_t(v, B)   \biggr\} _ { 1 \le t \le T} 
\\
& := \Vb (B) +  \sum_{v  \in B }\beta_{v, B} \tb  ,   
\stepcounter{equation}\tag{\theequation}\label{eq:ZB}
\end{align*} 
 { Moreover, for a given community $B$ that is significantly connected across time $t = 1,..., T$, we write the vector of node-set connectivity $ \Zb(B) $   as the sum of \textit{trend} and \textit{variation} components, where  $\beta_{ u,B} \tb$  represents the  trend component   which is linearly dependent on time and  $\Vb (B) $ represents the variation component that is stationary across time. }

The previous sections describe discovery of node-sets that are significantly connected across time, this section details testing for their trends. 
If the trends are posited to be  positive  or negative, then one-sided t-tests are used, respectively with null hypotheses  $H_{0,+}:  \beta_{ v, B}  \le 0  $ and  $H_{0,-}:  \beta_{ v, B}  \ge 0  $.
More detail can be found in the appendix (section \ref{sec:appendix}). If the the trend is positied to be negligible (stable), then the two sided test:
\begin{align*} 
H_0:&  \beta_{v,B}\ne 0  ; \quad \quad H_1:  \beta_{v,B} = 0 
\end{align*} 
is used. The hypothesis is flipped (compared to the positive or negative tests) in order to test if the  trend is equal to zero. We invoke equivalence testing methods (\cite{dixon_trend_2008}) to determine significance in relation to a pre-selected  symmetric interval $[-U,U]$ about zero.

Given an set $B_k$ at iteration $k$,  we first  find all nodes  $v^*$ that are  \textit{significantly bordering across time} as described in Section \ref{sec:significant_nodes} and label these nodes as $M_k(B_k)$ as in \eqref{eq:M_def}. 
We then  assess the significance of the trends of each of the nodes $v \in M_k(B_k)$ in relation to set $B_k$. Calculation of trend employs  test statistic for node-set connectivity $S(u,B_k, {G}_t)$:  
\begin{equation} \label{eq:NormalizeduB_t}
\Zb (v,B_k) =  \biggl\{  \frac{ S(v,B_k, {G}_t) - \Expec[ S (v,B_k, \mathcal{G}_t) ]  }{\Var( S(v,B_k, \mathcal{G}_t)) }  \biggr\} _ {  1 \le t \le T}. 
\end{equation} 
 
Using  $B_k$ and $ M_k(B_k)$, we then find the time trend  $\beta_{v,B_k}$  for each $v \in M_k(B_k)$. 
We assume that intertemporal communities have trends that are \textit{increasing},  \textit{decreasing}, or \textit{neutral}. 
We use the equivalence testing method to assess trend significance \citep{Schuirmann1987, dixon_trend_2008} . Even if a trend is significant, its impact may be negligible and should be assumed to be ``zero".
A bounding energy barrier $U>0$ is chosen to control the size of the desired time-trends.
A positive $U$ is chosen as a lower bound for a positive trend,  
$-U$ is  used as a upper bound for a negative trend. A symmetric bounding interval of $[-U, U]$ about zero is used for a neutral trend.

Hypothesis tests are conducted for the time trend  for set $B_k$ (at iteration $k$) and node $v$. 
Significances of  trend $\beta_{v,B}$ (assuming fixed  $B:=B_k$ at iteration $k$) are calculated using the difference of the estimates with the upper  bounds $U$ (if positive) and lower bound $-U$ (if negative). 
T-tests for these differences $\beta_{v,B} - U $ or   $\beta_{v,B} + U $ are then performed to assess significance while excluding very small trends. To determine whether a node-set has a  significantly negligible (neutral) trend, we utilize the approach outlined by Dixon et al. \cite{dixon_trend_2008} and use two one-sided tests to determine if $\beta_{v,B}$ is significantly outside the interval $[-U,U]$. 

Details on the test statistics can be found in the appendix in \ref{app:inc_dec_trend} for increasing and decreasing trends and \ref{app:neu_trend} for the neutral trend.
Before iterative testing in the general case for step $k$,   we first initialize according to section  \ref{app:init_TS}. 

\subsection{Iteration and Overlap Filtering Steps}

 After the procedures for selecting nodes that are both  significant in connectivity (Section \ref{sec:significant_nodes}) and  trend $\beta_{v, B} $  depending on the posited direction of trajectory (Section \ref{sec:trend_test}),  we derive p-values from the $t$-statistic of the time-trend . The nodes whose trends are significant after  incorporating the FDR  correction with significance level $\alpha$, are retained.
 
We update the set $B_{k+1}$ with the inclusion of the new nodes $v$ that are both significantly connected to $B_k$ across all time $t$ and have a significant trend according to the trend hypothesis. The procedure is repeated until the set becomes stable such that $B_{k} = B_{k+1}$ for all candidate sets. In all the applications used in this study, this process takes 3 to 5 iterations. 

After stable sets are found from the iteration steps, they are filtered by their Jaccard overlaps \cite{palowitch_continuous, He2019}. We use an overlap threshold of 0.50 to remove clusters with over 50\% overlap; more details on this procedure can be found in prior work \cite{palowitch_continuous}. After filtering by Jaccard overlaps, communities of size 3 or less are removed, as dyadic or triadic  relationships between nodes may be too localized to be meaningful in a larger scale.

 \subsection{Effect of Normalizing Edges} \label{sec:normalizing_edges}

\begin{figure} [htbp]
	\centering	
	\textbf{Global Variance $\kappa_t$ (Divvy; Citibike; NYC Taxicab)}
	\\
	\boxed{
	\includegraphics[width=0.31\linewidth,  	trim={1cm 0cm 1cm 1.8cm},clip]{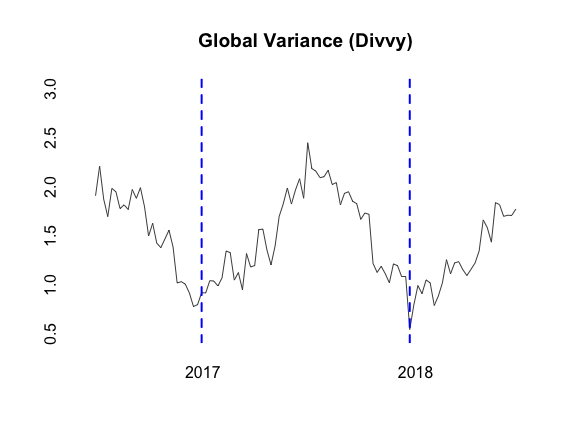}
	\includegraphics[width=0.31\linewidth, 	trim={1cm 0cm 1cm 1.8cm},clip]{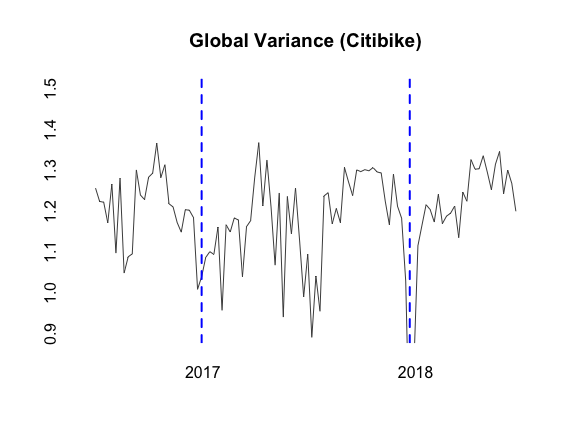}
	\includegraphics[width=0.31\linewidth, 	trim={1cm 0cm 1cm 1.8cm},clip ]{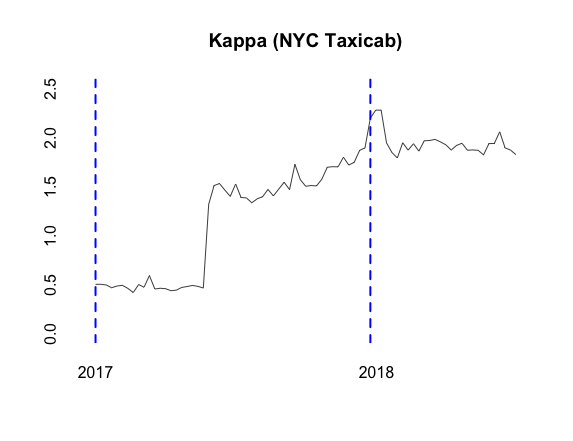}  
}
	\caption{ Global variance parameter $\kappa_t$ from 2016 to 2018 for the Divvy system in Chicago (left), the Citibike system in New York City (center), and $\kappa_t$ for NYC taxicab networks (right) from 2017 to 2018 }
	\label{fig:kappaTS}
\end{figure}

Modeling network time-series using the weighted configuration model places edge-weights in a relative scale when they are normalized by their expectations and variances, which are functions of  {global $\kappa_t$. 
Global $\kappa_t$ is shown} to be highly seasonal (fig. \ref{fig:kappaTS}) in the Divvy system in Chicago but less so for the NYC taxicab and Citibike data. The variances in the taxicab data \jf{experience} a sudden increase in the middle of 2017 and \jf{thereafter consistently increase} through time. The high seasonality of $\kappa_t$ in Chicago and the effects of its removal by normalization \jf{are} apparent in figure \ref{fig:example_SS}.
Scaling edge weights is especially useful in time-series networks where seasonal effects dominate much of the variation (in the Divvy data) or the trend (in NYC taxicab data).

\section{ Results}\label{subsec:primary_findings}
\ch{We report results for a range of values for tuning parameters $\alpha$ and $U$ for observed demand $\{G_t\}_{1\le t \le T}$ . 
}

In the Divvy Network, we fix $\alpha$ at 0.05 and $U = 0.007$ as well as $0.009$ because these settings \jf{capture} clusters of moderate sizes across all trend categories and also \jf{show} distinct geographical divisions. Under these tuning parameters, we \jf{find} five clusters with decreasing connectivities over time  and five clusters with increasing connectivities. \jf{We find only one cluster with a stable trend} at the 0.05 significance level.

\begin{figure} [htbp]  	
	\centering 	
			\includegraphics  	[width=0.85\linewidth, trim={0cm  7cm  0cm 0cm, clip} ] {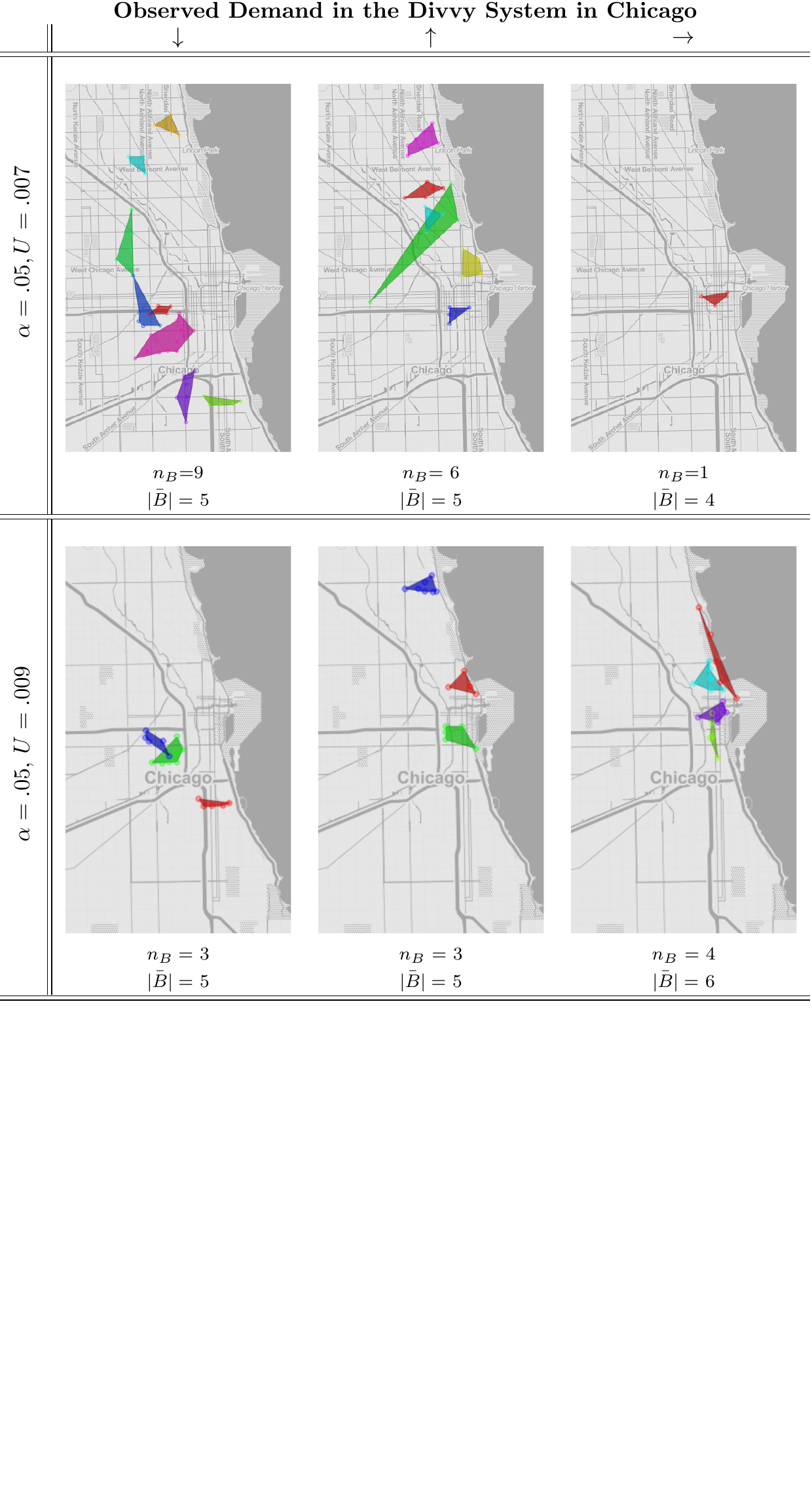}  
	\caption{  	\footnotesize{}	 Intertemporal communities of increasing or decreasing trends amongst Divvy stations in 2016-2018 under varying significance levels  and bounding parameters $U$ using the network time-series $\{G_t\}$ uncorrected for load-imbalance.  $n_B$ represents the number of found communities and $\bar{|B|}$ represent the mean size of communities.  }\label{fig:Obs_Chi}
\end{figure}

There is a stark division in trends between the northern and southern parts of the city (fig. \ref{fig:Obs_Chi}).  At the 5\% significance level, clusters with significantly decreasing trends \jf{are} mostly found in the southern and western parts of the city, while clusters with significantly increasing trends \jf{are} mostly found in the northern and central parts of the city. Interestingly, the decreasing clusters map to a nearly concentric outer ring around the central part of the city, while the increasing clusters stretch from the Loop northwards along the shore of Lake Michigan. One stable cluster is located in the Loop.

\begin{figure} [htbp]
	\centering	 	
	\textbf{Sum of Edge Weights in Sample Community}	
	\includegraphics[width=0.65\linewidth, trim={0cm 0cm 0cm 2cm},clip]{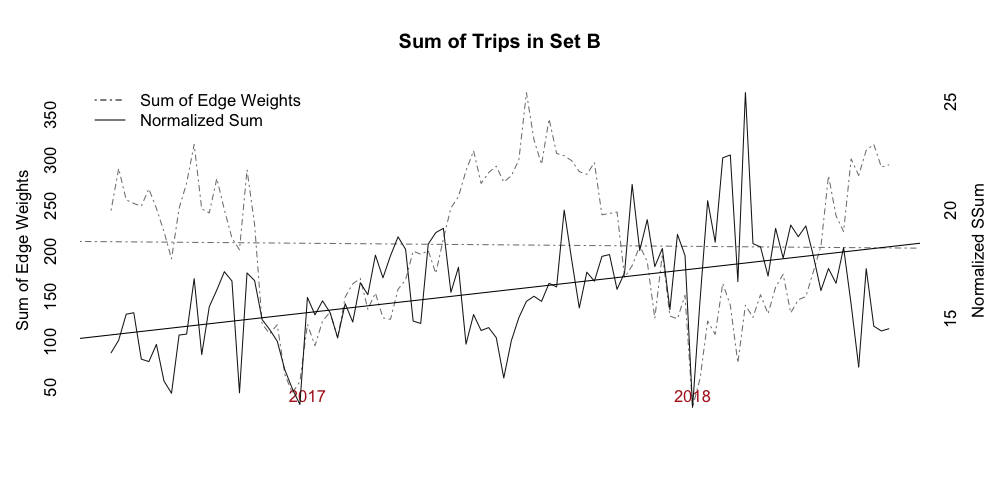}\centering  
	
	\boxed{\includegraphics[width=0.6\linewidth,trim={4cm 0cm 0cm 0cm},clip]{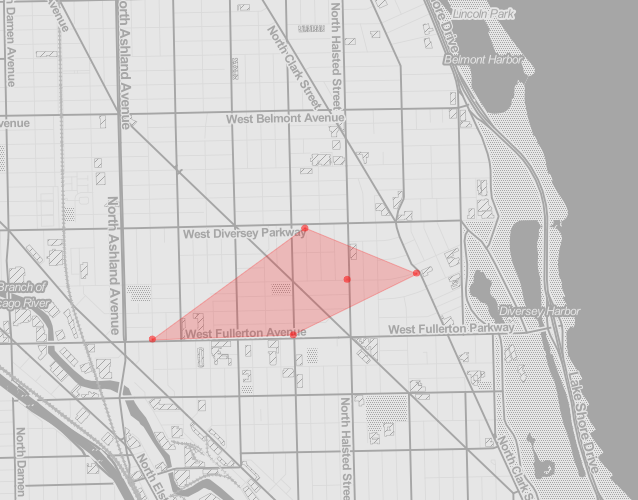} } 
	
	\caption{ \footnotesize{} \textit{top}: Total trips in a community in networks $ G_t $ with increasing normalized connectivity over time comprising 5 stations around the {Lincoln Park}  Neighborhood in Chicago. \textit{bottom:} Map of stations in $B$.}
	\label{fig:fig1tsplot}	  \label{fig:example_SS}
	
\end{figure}

It is useful to focus on one community to illustrate the effect of edge normalization (see section \ref{sec:significant_nodes}). In figure \ref{fig:example_SS}, while the raw edge weights show a stable trend, the normalization $\Zb(B)$ shows an increasing trend. Thus, the collection of five stations in the Lincoln Park neighborhood in Chicago is classified as a cluster with an increasing time trend rather than a stable one.

\begin{figure} [htbp]  	
\centering 	
		\includegraphics  	[width=0.85\linewidth, trim={0cm  7cm  0cm 0cm, clip} ] {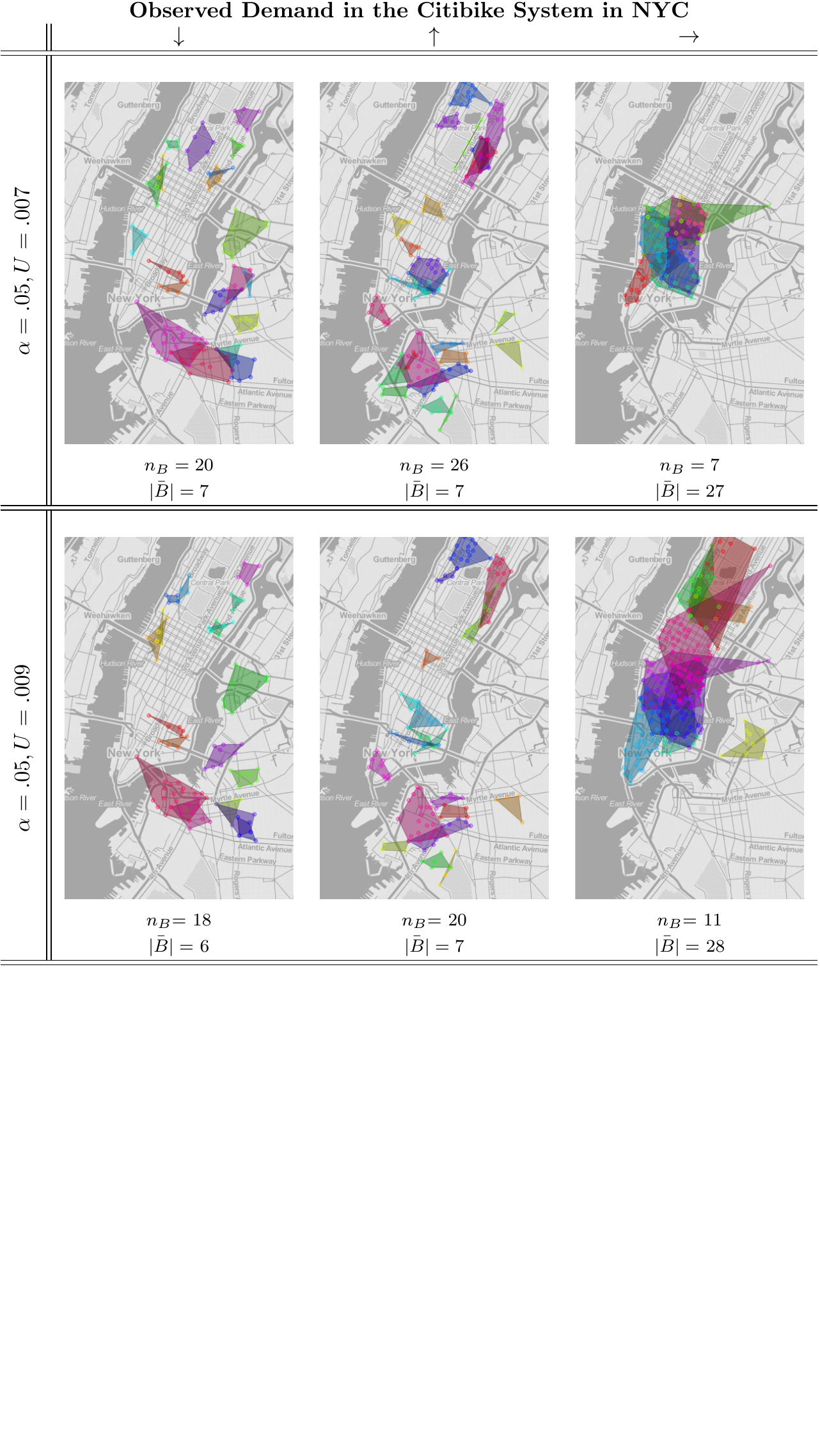}  
	\caption{Intertemporal Communities of increasing ($\uparrow$), decreasing ($\downarrow$), and stable ($\rightarrow$) trends amongst stations in years 2016-2018 under varying significance levels  and bounding parameters $U$ in the uncorrected networks $G_t$. $n_B$ represents the number of found communities and $\bar{|B|}$ represents the mean size of communities.   }
	\label{fig:Obs_NYC}
\end{figure}

In New York City, many clusters are found in the Citibike system in networks of observed demand (raw counts of trips) $G_t$ (fig. \ref{fig:Obs_NYC}).   We set $\alpha$ to 0.05 and $U$ to 0.007 and 0.009 to allow direct comparison to the clusters in the Divvy system in Chicago. When $U$ is set to 0.007, increasing and decreasing clusters are found throughout Manhattan and Brooklyn while stable clusters are mostly concentrated around Lower Manhattan.  When $U$ is set to 0.009, increasing and decreasing clusters decrease in size and number but stable clusters increase in size and geographical scale: All of the land-area in Manhattan and part of north Brooklyn is covered in these stable clusters. 

The geographical domain that the taxicab network covers is much larger than the bikeshare network, which only spans Manhattan and Brooklyn. 
\jf{In two settings of $U$, clusters are  decreasing in connectivity across much of the Bronx, Queens, and much of Brooklyn.} 
Clusters are consistently increasing in eastern parts of \jf{Queens. One cluster} appears to consistently link Staten Island to southern Brooklyn for both values of  $U$.
\jf{Clusters are stable around the denser parts of the city, as is the case in Upper Manhattan when $U$ is 0.01} and \fixit{in Upper and Lower Manhattan, Central Brooklyn, and Astoria in Queens  when $U$ is 0.02.}

\begin{figure} [htbp]  		
	 	\centering 	
	 \includegraphics	[width=0.85\linewidth, trim={0cm  10cm  0cm 0cm, clip} ] {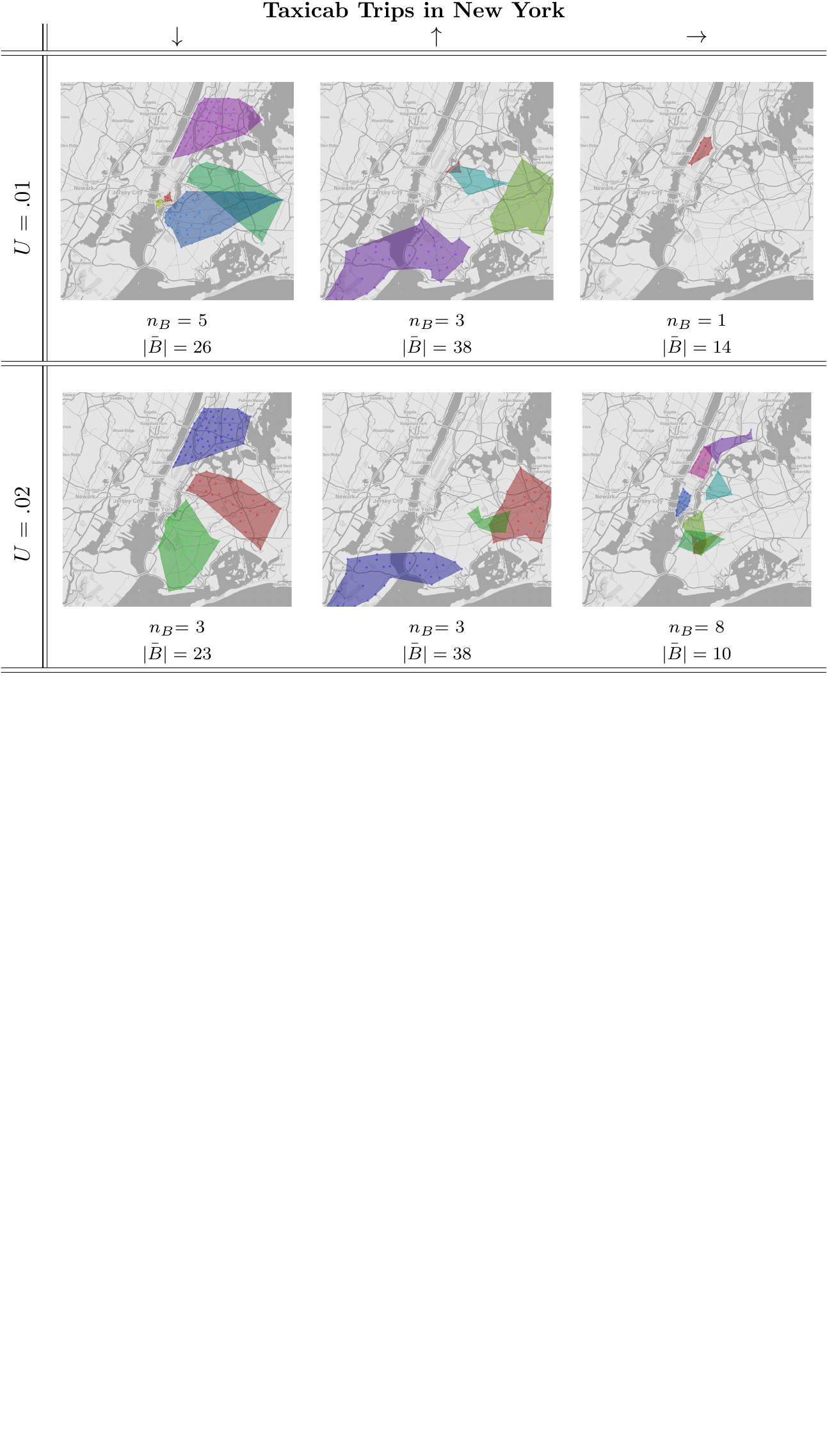}  
	\caption{Intertemporal Communities of increasing, decreasing, and stable trends in taxicab networks amongst zones in years 2017-2018  in New York City under varying significance levels  and bounding parameters $U$. $n_B$ represents the number of found communities and $\bar{|B|}$ represents the mean size of communities rounded to the nearest integer.}
	\label{fig:Taxi_NYC}
\end{figure}

\ch{\subsection{Effect of Demand Correction}}

We apply the intertemporal community detection algorithm \jf{to the} demand-corrected (DC) time-series  networks $ \{ \tilde{G}_t \}_{1\le t \le T}$ with weights $\tilde{W}_{uv,t}$ in the Citibike system.  We use the same significance $\alpha=0.05$ and set barrier $U$	to 0.007 and 0.009 as in observed trip networks in NYC and the Divvy system in Chicago. The obtained communities retain similar geographical characteristics  as those in uncorrected graphs, but with some key differences.

When $U$ is set at 0.007,  the decreasing and increasing clusters in the demand-corrected networks are localized in approximately similar geographical regions as in non-corrected networks. Increasing clusters are mostly located in Upper and Lower Manhattan as well as Southern Brooklyn. Decreasing clusters are present in some small areas throughout Manhattan but pervasively cover swathes of northern Brooklyn around the Williamsburg region. Stable clusters mostly span Midtown Manhattan but also extend to northern Manhattan and parts of Brooklyn.

\begin{figure} [htbp]  	
	\centering 	
\includegraphics  	[width=0.85\linewidth, trim={0cm  8cm  0cm 0cm, clip} ] {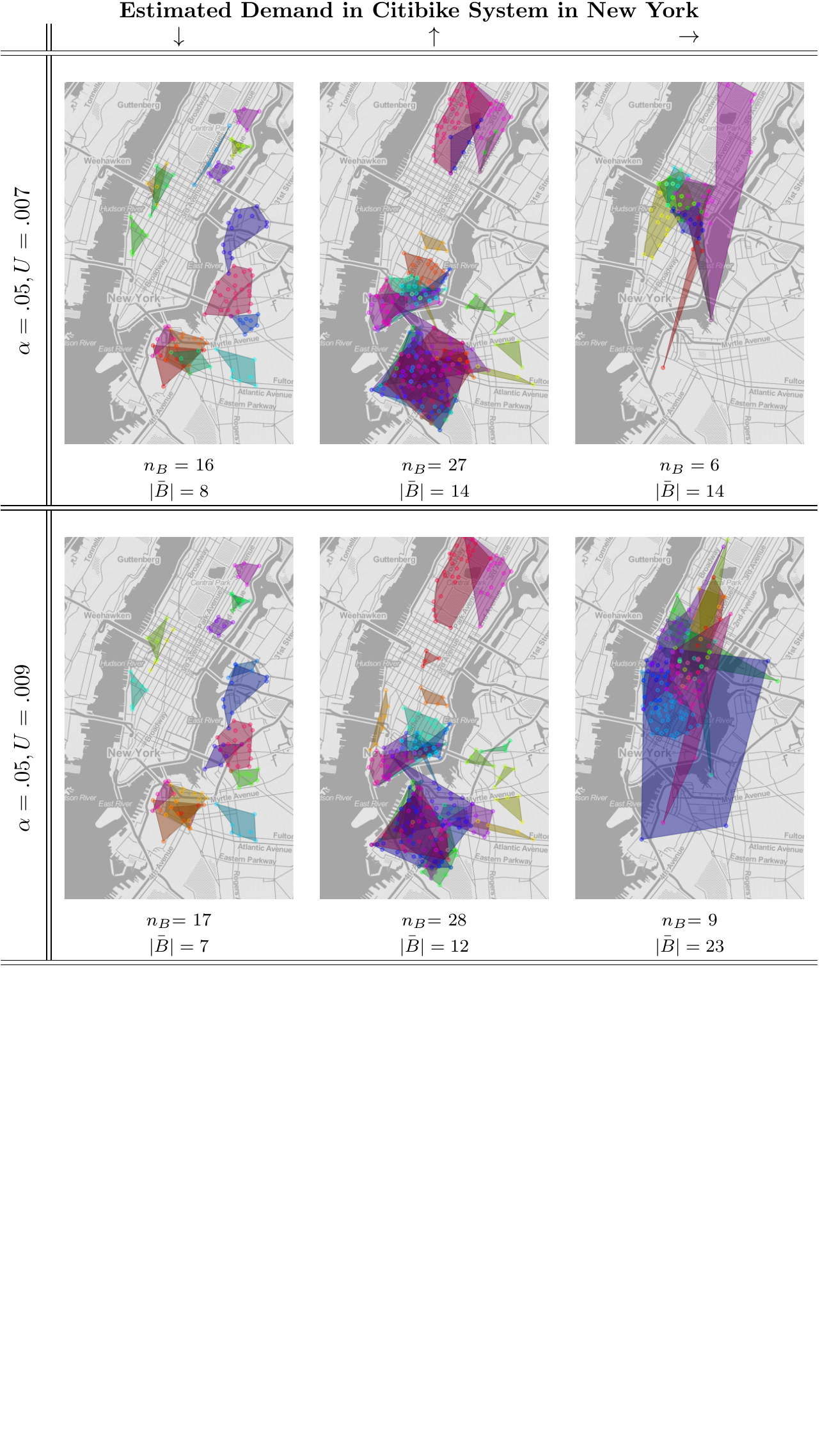}  
	\caption{Intertemporal Communities of increasing, decreasing, and neutral trends amongst Citibike stations in years 2016-2018  in New York City under varying significance levels  and bounding parameters $U$ in the demand-corrected networks $\tilde{G}_t$. $n_B$ represents the number of found communities and $\bar{|B|}$ represents the mean size of communities rounded to the nearest integer.}
	\label{fig:Adj_NYC}
\end{figure}

When $U$ is increased to 0.009, the increasing and decreasing clusters shrink in size and number and the stable clusters expand. Increasing clusters are more visibly located in Upper and Lower Manhattan (similar to the clusters in the graphs of observed demand) at the higher threshold. Decreasing clusters are interspersed throughout the city but large coherent areas are more clearly located around northern Brooklyn, also as in the observed graphs $G_t$. The stable graphs, however, are much larger and cover much more ground in Lower Manhattan (fig. \ref{fig:Adj_NYC}) .

\section{Discussion}

\chsb{In the Citibike, Divvy, and NYC taxicab systems, we observe a trade-off between increasing or decreasing clusters and stable clusters \jf{depending} on the choice of $U$. If $U$ is larger, then there is ``more room" for a trend to be classified as stable, but less so for increasing or decreasing trends.}  Discovery of more increasing and decreasing clusters when $U$ is increased suggests that these clusters are increasing or decreasing in connectivity \jf{at different} rates from the other clusters. When $U$ is large,  increasing and decreasing clusters vanish but more stable clusters persist.

The interaction between $\alpha$ and $U$ is not entirely linear or monotonic. Though a decrease in $\alpha$ may correspond to an increase in $U$, a lower $\alpha$ implies that the nodes are more connected at each time-instance and does not necessarily mean that the trend is higher.   
Figures \ref{fig:Obs_NYC} and \ref{fig:Adj_NYC} show that in both $G_t$ and $\tilde{G}_t$,  clusters appear as $U$ becomes larger and $\alpha$ stays the same. Such behavior may be attributed to FDR correction. 
A lower barrier $U$ may yield more  significantly connected nodes  but with weaker trends. The sensitivity of community detection to the choice of parameter is an important issue \citep{austwick_structure_2013}.  We compare the extracted communities under different tuning parameters $U$ and $\alpha$.


In results from the observed network $ G_t $ in Chicago,  the choices of $\alpha$ and $U$ produce generally similar results over a range of values (fig. \ref{fig:Obs_Chi}). Shifting $U$ from $.007$ to $.009$ induces discovery of more increasing and decreasing clusters, but the bound is too tight for any  significant sets \jf{to be found} under the hypothesis tests in \eqref{eq:neutral_trendhypothesis_ab}.  

 { More increasing clusters are found in high-traffic areas surrounding cultural amenities such as the Adler Planetarium in the Loop when $U=0.005$ .  Several other clusters in the Near North Side and the Loop \jf{are} found under this threshold that  {are} not  {found} under the settings in figure \ref{fig:Obs_Chi}. Stable clusters are mostly found also in the Loop and the Near North Side and  share a lot of commonalities with increasing clusters at less stringent thresholds; we deduce they have a moderate increasing trend.}  

\ch{ Our analysis is exploratory in nature and only summarizes the trajectories of network structures in time but not their underlying causes. While this work is focused on methodological aspects of temporal community detection, results suggest that it might be useful to think about the causal mechanisms that underlie the different types of clusters in the bikeshare networks. The geographical patterns of the cluster map to different neighborhood characteristics.}

\ch{ Chicago is an conventionally viewed as a monocentric city focused on the downtown \cite{10.1371/journal.pone.0137922}\jf{,} so it is unsurprising that stable clusters are found around the Loop. The southwestern part of the city is comparatively less affluent. Pilsen, in particular, is a predominantly Latino neighborhood, although its demographic composition is rapidly changing due to gentrification  \citep{betancur_traj,hope6}.   Lincoln Park and Lakeview are known to be  affluent and mostly white residential neighborhoods. Increasing, stable, and decreasing trends map closely to these neighborhoods of differing socioeconomic characteristics and suggest latent factors undergirding the decreasing trends that are driving the overall slight decrease in trips from 2016 to 2018.}


At the same fixed parameters for $U$ and $\alpha$,  clusters in NYC are more numerous and less geographically spread out than Chicago, possibly because the city is much denser and more populous. Moreover, the seasons are milder, which induces less variation in trends. Figure \ref{fig:kappaTS} shows that the global variance parameter $\kappa_t$\jf{ of} the Divvy system is highly seasonal, \jf{unlike that of} the Citibike system.


Several areas appear to be persistently decreasing in both demand-corrected and observed networks. The easternmost group of four stations in \textit{Williamsburg} is found in both thresholds of $U$ in demand-corrected and observed networks, suggesting a plausibly  {real} relative decreasing trend in this neighborhood.

Citibike  has much higher overall usage and thus a lower significance level $\alpha$ (compared to Divvy) is used to locate the clusters that are significantly anomalously connected. The Citibike system is globally consistently increasing as opposed to Divvy which is slightly decreasing from 2017 to 2018.


The edge-normalizing step of the community detection algorithm (section \ref{sec:normalizing_edges}) makes such station-specific adjustments affect the whole network, thereby affecting the entire system. Regardless, similarities persist in clusters in both DC and uncorrected graphs. The demand corrected (DC) networks $\tilde{G}_t$ when $U$ is 0.007 and 0.009 yield similar decreasing clusters to \jf{those of the} uncorrected networks $G_t$. 

Demand-adjustment makes a considerable difference in some clusters in the Citibike system. Among decreasing clusters, most of the discovered communities in both uncorrected and corrected graphs are located in Brooklyn, with some scattered around Manhattan (figs. \ref{fig:Obs_NYC}, \ref{fig:Adj_NYC}). However, clusters in general are smaller  and less numerous in the DC case. The opposite case is observed for increasing clusters: discovered communities in DC graphs $\tilde{G}_t$ are much more geographically defined in  Upper and Lower Manhattan and southern Brooklyn and notably much larger, more numerous, and overlapping. Stable clusters are confined to Manhattan in the observed graph $G_t$, but appear to form different shapes and extend to Brooklyn in DC graphs $\tilde{G}_t$.

Adjusting for demand-correction thus reveals stronger, more cohesive increasing trends \jf{within the} ridership and  {suggests} that observed trips do not adequately capture the latent increasing signals that are distorted by load imbalances from empty stations.

A common feature of  decreasing clusters in both DC and non-DC networks across all choices of $U$ is the  presence of large clusters around the Williamsburg neighborhood in north Brooklyn. Small pockets of the neighborhood are clustered into increasing or neutral clusters, but decreasing clusters are dominant.  On the other hand, south Brooklyn has more of a mixture of trends. Though certain regions in DC and non-DC graphs are grouped into decreasing clusters, the decreasing clusters in DC graphs are very large, encompassing nearly all of the southern part of the city,  {with some clusters extending} to southern Manhattan.

Due to the fact that trips by taxicabs are, in general, longer than the trips by bike, the clusters that are found cover larger  {areas.} 
We choose $U$ to be 0.01 and 0.02 to maximize differentiation between clusters based on  {an} assumed negligibility of trend. 
Most increasing clusters generally look the same with one exception. One of the decreasing clusters \jf{changes} shape rather drastically as $U$ is \jf{increased} from 0.01  to 0.02\jf{ and goes from} covering nearly all of Brooklyn to covering northern Brooklyn and part of Queens. These patterns mostly demarcate general regions and may illustrate decline in usage within these areas due to other transit options such as bikeshares in Brooklyn and the Bronx. Increasing clusters are generally similar across choices of $U$ and highlight the corridor between Brooklyn and Staten Island, which may be illustrative of the lack of public transit between these regions. The neutral clusters only demarcate a small region around Central Manhattan when $U$ is 0.01, but expand to cover most of Manhattan when $U$ is 0.02. These patterns may be indicative of the consistent usage of taxicabs in the busiest parts of the city. \chsb{There are some similarities between clusters of taxicab trips and Citibike trips: stable clusters are found around Central Manhattan in both cases (DC and non-DC cases for Citibike), and southern Brooklyn is found to be increasing (DC case) just as in the taxicab network (though the latter case is connected to Staten Island)} However, because the geographical scale of the taxicab dataset is much larger than that of Citibike, and signal different kinds of mobility and connectivity, the similarities in clustering geographies may not be reflective of similar underlying trends.

Results in Chicago and New York illustrate the similarities and differences between the two systems. The Citibike system is larger than Divvy (18 million trips in 2018 vs 3.6 million) and is growing at a much faster rate. As such, the significance thresholds for connectivity are stronger for Citibike and \jf{its} communities represent more densely connected collections of nodes. \jf{There are no} large contiguous areas of consistent decreasing trends like in the ring of clusters surrounding Chicago's city center (fig. \ref{fig:Obs_Chi}). 
The decreasing clusters in Chicago may be {indicative} of specific geographical patterns that  drive the global slight decrease in the Divvy system  because the clusters are geographically coherent. However, the decreasing clusters in NYC  are distinctively countervailing  {with respect to} to the overall increasing global trend, though certain clusters (like in Williamsburg) are persistent across several settings and parameters. 

Because the proposed method is  for exploratory purposes, these summarizing claims  should be verified in a more  rigorous way in future research. Furthermore, the choice of $U$ varies by application. We use an ad-hoc scheme to select  $U$  whose resultant neutral clusters yield approximately the same amount of nodes as the increasing and decreasing clusters combined. 

However, because  most of the results we present include two different values of $U$ in order to show the differences in results due to adjusting the parameters , the results in this study may not strictly adhere to this criteria. However, results from Fig. \ref{fig:Obs_Chi}, Fig. \ref{fig:Obs_NYC} and  \ref{fig:Adj_NYC}  all approximately follow ths heuristic  when $U=.009$, though they may have different $\alpha$'s. Different applications of intertemporal community detection may call for different criteria for tuning parameters. For example, setting $U$ to be small so as to not allow discovery of any neutral clusters (i.e. Fig. \ref{fig:Obs_Chi}, $U=.007$) may also be a suitable option.
In future work, more principled approaches for setting tuning parameters utilizing cross-validations may be investigated.


\section{Future Work}

 In both Chicago and New York City, there may be several explanations for the  underlying signals that cause the clusters to decrease in connectivity.  
Further work may examine what these signals are and how these signals may function.
One explanation may be that decreasing trends are symptoms of  {displacement,} destabilizing  steady ridership  among long-term inhabitants in gentrifying neighborhoods.   Another may be differential rates of attention given to load rebalancing in stations in different neighborhoods with varying resources. Causal analysis of these phenomena are outside the scope of this study, but our exploratory results are useful in initializing conversations about changes in mobility patterns \jf{within and between} neighborhoods.  Future work may analyze the relationship between the discovered communities and factors such as new construction, bike lanes, weather, incomes, and demographic characteristics.

The methods devised in this study can be applied to a variety of data in network time-series format, particularly human mobility networks.   { The method can be applied to bikeshare networks in other cities, or may be applied to other networks of transportation in urban systems.}
Future work may elaborate on the theoretical properties of intertemporal community detection. The  null model described in section \ref{sec:intertemporal_null} may also have further use in statistical inference or \jf{in} forecasting  future patterns. Another  extension \jf{would be} to account directly for the spatiotemporal aspects of trips in the methodology.  

  Our work currently relies on historic station inventory data for the analysis of the Citibike system. We do not have access to historical inventory data for Chicago and thus are not able to estimate demand. Though similarities between corrected and non-corrected networks in the NYC bikeshare system shows that there may be some use in using only non-corrected data in Chicago, there are limitations in drawing conclusions for demarcations of functional mobility zones using only observed demand.
  We propose a method in \ref{app:demand_correction_est}, but further  estimation of demand without historical station inventory data should be explored in future work in conjunction with community detection.

\section{Conclusions}

\chsb{We proposed a novel method to cluster  networks representing bikeshare systems that vary across time. Our community detection method combines usage of a configuration null model  with \jf{a} trend  model to describe the expected trajectory of the graph evolutions. We use a significance-testing methodology to assess whether nodes are anomalously connected to each other within and across time-periods.}  By using the proposed method, we are able to filter some of the system-wide seasonal effects and \ch{map geographically coherent communities} of latent human mobility signals in the bikeshare stations in Chicago and New York and the taxicab network in New York. The methods used in this paper may be applied to other situations where it is important to study the evolution of structures within networks.

\appendix
 \section{Appendix} \label{sec:appendix}
 \subsection{Corrections for Forgone Trips Due to Load Imbalance Given Load Rebalancing Data} \label{sec:demand_correction}

We estimate the functionals $ \Prob( \tilde{E}_{u,Y}) $ by taking the average rate \jf{at which} a station \jf{is} empty i.e. yields no available bikes. $ \Prob( \tilde{E}_{u,Y}) $  are calculated as the ratio of the time-intervals that a station is empty to the total intervals during peak-times (i.e. when users could plausibly check out or return bikes). The ratio represents the probability of a station being empty when a user accesses it. A high ratio signifies that the station is usually empty, and so it is more frequently load-imbalanced due to high usage, hence more weight should be proportionally accounted for to estimate the trips that could have been taken if the system was perfectly balanced.  

Let $\tilde{E}_{u,Y}$  be the event that a typical trip in year $Y$ from or to station $u$ is foregone owing to load imbalance and let  $\Prob( \tilde{E}_{u,Y})$ be its associated probability. $\Prob( \tilde{E}_{u,Y})$ is approximated as:  
\begin{align} \label{eq:Pforegone}
\Prob( \tilde{E}_{u,Y})  
\approx   \frac{     \#   \{ \text{intervals when  } u \text{ is empty in year  } Y\}}{  \#  \{ \text{total intervals in station } u \text{ in year } Y \} }.
\end{align}

{Observed demand} $W_{uv,t}$ for each edge between stations $u,v$ during time-index $t$ (weeks in this analysis) are then converted to {estimated demand}  $\tilde{W}_{uv,t}$  as follows:
\begin{align*}
\tilde{W}_{uv, t} = W_{uv, t } ( 1+  \Prob( \tilde{E}_{u,Y}) ) ( 1+  \Prob( \tilde{E}_{v,Y}) ), \quad  t \in Y. 
\end{align*}

We refer to the time-series of graphs comprised of these demand-corrected (DC) weights as $\{\tilde{  G }_t \}_{ 1 \le t \le T}$. For this study, we assume that this probability is constant over the year. Seasonal effects may be influential in this calculation but will be deferred to future research.   We assume that a full station induces a negligible impact on load imbalance compared to empty stations. Each probability is  calculated as the proportion of time-intervals that the station is empty. We construct networks of estimated demand to correct for trips that could not have taken place due to full or empty stations and find communities within these networks to more accurately find communities of trip demand in a human mobility network \cite{FAGHIHIMANI201553,Liu2016RebalancingBS,Liu2016RebalancingBS}.

 \subsection{Corrections for Forgone Trips Due to Load Imbalance Without Rebalancing Data}
	\label{app:demand_correction_est}

Though real-time data on station status (e.g. number of open slots) exist and are available online  \cite{divvy_gfbs}, we do not have access to the historical load rebalancing data and as such we need to estimate the probability of foregone trips. To determine the presence of these forgone trips induced by full or empty stations, we look for anomalous gaps in usage of stations on the days that it is heavily utilized. We refer to these gaps due to forgone trips as \textit{load imbalance}.  We describe a simple significance-testing based method that corrects the  counts of trips between stations (edge weights) in each graph $G_{t}$ for week $t$ in each year $Y$. We have omitted the results of this analysis of the Divvy System in Chicago, though results from this study can be made available on request.

Let $\tilde{E}_{u,Y}$  be the event that a typical trip in year $Y$ from or to station $u$ is foregone owing to load imbalance, we write  $\Prob( \tilde{E}_{u,Y})$ as its associated probability. For this paper, we assume that this probability is constant over the year. Seasonal effects may be influential in this calculation but will be deferred to future research.  

Sums-of-trips $W_{uv,t}$, or \textit{observed demand}, for each edge between stations $u,v$ during time-index $t$ (weeks in this analysis) are then converted to \textbf{estimated demand}  $\tilde{W}_{uv,t}$  as follows 

\begin{align*}
\tilde{W}_{uv, t} = W_{uv, t } ( 1+  \Prob( \tilde{E}_{u,Y}) ) ( 1+  \Prob( \tilde{E}_{v,Y}) ), \quad  t \in Y 
\end{align*}

We refer to the time-series of graphs comprised of these demand-corrected weights as $\{\tilde{  G }_t \}_{ 1 \le t \le T}$. We now describe how to estimate the functionals $ \Prob( \tilde{E}_{u,Y}) $.

\subsubsection{Calculating Significant Gaps in Station Activity}

A time interval for station $u$ is an interval between any two consecutive events (arrivals or departures). We first formulate a methodology to judge if a time interval is anomalous or not. We call such  an unnaturally long time interval a gap. Gaps may occur because of load imbalance or random events not related to load imbalance. We posit that the probability of the occurrence of a foregone trip is:

\begin{align} \label{eq:Pforegone}
\Prob( \tilde{E}_{u,Y})  
\approx   \frac{     \#   \{ \text{gaps in station  } u \text{ in year } Y  \text{ due to load imbalance} \}   }{  \#  \{ \text{intervals between trips in station } u \text{ in year } Y \} }.
\end{align}

We assume that typical waiting times (in seconds) between consecutive events (start and end of trips) at a station $u$ on day $d$, $w_{u,d}$ follows an exponential distribution with mean $\delta_{u,d}$  \cite{Gast:2015:PFB:2806416.2806569}.  Note that the cardinality of waiting times is equivalent to the strengths $S_{u,d}$, or sum-of-trips, of station $u$ on day $d$ subtracted by 1.  We filter out the first and last 10\% of trips that occurred during day $d$ are censored to filter out the longer gaps during the early and late times of the day, hence only restricting the times $s$ to non-dormant hours, so let $S^*_{u,d}-1$ represent the number of trips excluding  the first and last $10\%$ of trips.  We count the number of anomalies\textit{ per day} assuming that high-activity stations are rebalancing at least several times a day \cite{pendem2019}. To determine anomalies in durations between activity, we first define waiting-times. Let $\theta_{1,u,d} < \theta_{2,u,d} <...<\theta_{S^*_u,u,d} $ denote the time points of consecutive activity on day $d$ at station $u$ after removing the upper and lower $10\%$ of trip-times.

Let $\mathcal{S}^*_{u,d}$ represent the collection of intervals $  \{[ \theta_{i,u,d} , \theta_{i-1,u,d} ] \}$  and let $w_{i,u,d} = \theta_{i,u,d} - \theta_{i-1,u,d} $ denote the length of these corresponding intervals. We define the   sample mean $\bar{ \delta}_{u,d}$ as
\begin{align*}
\bar{ \delta}_{u,d} = \frac{1}{S^*_{u,d}-1} \sum_{i = 1}^{S^*_{u,d}}  w_{i,u,d}.  
\end{align*}


Let  $I_{u,d}$ be the number of time-intervals  $w_{u,u,d} \in \mathcal{S}^*_{u,d}$ whose lengths are significantly greater than  $ \delta_{u,d} $ under significance level $\alpha$ after being corrected by  the Benjamini-Hochberg false-discovery rate rejection procedure \cite{bh_reject}. This procedure will be described in the later section \ref{sec: ordinary_BH} and will be used in the community detection algorithm. Precisely:

\begin{align*}
I_{u,d} =
\# \{  w_{i,u,d} : w_{i,u,d} >  \bar{ \delta}_{u,d}   \text{ at } \alpha, \text{ FDR corrected across } w_{i,u,d} \in \mathcal{S}^*_{u,d}  \}. 
\end{align*}

$I_{u,d}$ represents the estimated number of gaps in waiting-times. These values may represent gaps due to either load imbalance or typical events such as a break in usage during lunch, or an adverse weather event. We assume that these typical events are different from load  imbalance. We do not have data on events that could have led to these gaps caused by typical events. However, we can determine a summary measure of the gaps that occurred when the station is operating  \textit{in excess}, which we define as the condition when the number of trips is significantly greater than the number of slots in the stations. We can also determine the total sum of the gaps that may be due to random, \textit{typical}, conditions when the station is not operating in excess.  We posit that the difference of the gaps under these two conditions provides a reasonable approximation of the gaps owing to load imbalance.

 \subsubsection{Finding Stations with Excess Demand}

We define $C_{u,Y}$ as the carrying capacity, or number of slots, in a station $u$ in year $Y$. Typically, carrying capacities of stations are updated once per year. If    $C_{u,Y}$ of a station (in and outflows) are exceeded significantly at a given day $d$ by the total trips (daily strengths)  $S_{u,d}$, then we consider the possibility of a overfilled or empty station may influence the decisions of a potential user. We define \textit{excess demand} $D_{u^*,d}$ in stations $u^*$ where $\{u^*: S_{u^*,d}  \ge  C_{u^*,Y}\}$ as:

\begin{align}\label{def:capacity_Exceedance}
D_{u^*,d} = ( S_{u^*,d} - C_{u^*,Y} ) \sim \text{Poi} (\lambda_d)  
\end{align}  

We assume that the counts of  { excess demand} on day $d$ at station $u$ adheres to a Poisson distribution across all stations $u \in [ n]$ on day $d$. Functionals related to the total number of trips between periods of times are conventionally modeled as Poisson \cite{Gast:2015:PFB:2806416.2806569}. Let $\lambda_d$ be the typical network-level excess level of demand in day $d$ and let 
$\bar{\lambda}_d$ be its sample mean:
\begin{align*}
\bar{\lambda}_d = \frac{1}{n} \sum_{u = 1}^n D_{u,d}.
\end{align*} 

To determine whether station $u$ is operating \textit{in excess} on a given day $d$ in year $Y$, we use the Benjamini-Hochberg false-discovery rate correction ( section \ref{sec: ordinary_BH}) to find the stations that are significantly over capacity on day $d$. We evaluate the p-value of excess demand $D_{u,d}$ at station $u$ by  testing every $u \in [n]$ on day $d$ against the sample mean $\bar{\lambda}_d$ under a Poisson distribution under fixed significance $\alpha$.

We introduce a binary random variable $Q_{u,d}$ to denote if a station is significantly in excess. Let the value of  $Q_{u,d}=1$ if $ D_{u,d}$ is judged to be  significantly anomalous from $ \bar{\lambda}_d$ under significance level $\alpha$ with false discovery rate correction across stations $u^*$ with excess demand above 0, otherwise, let $Q_{u,d}=0$.  Note that $Q_{u,d}$ is zero for all $u$ such that  $\{u: S_{u,d} <  C_{u,Y}\}$, but it is zero for \textit{some} stations $u^*$ such that $\{u^*: S_{u^*,d} \ge  C_{u^*,Y}\}$.


 \subsubsection{Estimating Foregone Trips}

Gaps may be due to typical \textit{baseline} events or to load imbalance. On a given day, a station may be visited above or below its average rate of activity due to chance. However, if the station significantly exceeds demand (number of trips far exceed the number of slots) on such a day, then there is more reason to believe that the gaps in waiting-times between usage are plausibly related to load imbalance.We approximate the gaps using methods described in the previous sections. 

Let $\hat{g}_{u,Y}^E$ denote the total approximated number of gaps in activity in station $u$ over year $Y$ on the days $d$ when the station is operating in excess (i.e. $ Q_{u,d}$).  We assume that the indicator for station $u$  for a gap   is  {independent} of the fact that the station is over capacity on day $d$. The estimated counts of gaps when the station is operating in excess is expressed as:

\begin{align*}
\hat{g}^E_{u, Y}   & =        \sum_{d  \in Y } I_{u,d}   Q_{u,d}  
\end{align*}

Recall that $1-Q_{u,d}$ denotes the judgement by the FDR procedure of a non-anomalous demand on day $d$. Let \textit{$\hat{g}_{u,Y}^{b}$} denote  the sum of the number of gaps on days when the excess demand of  station $u$ is not significantly anomalous with respect to $\text{Poi} ( \bar{\lambda}_d) $. Here $ 1-Q_{u,d} =1$  representative of a typical day with \textit{baseline} anomalies. These counts are estimated as:

\begin{align*} 
\hat{g}_{u, Y}^b & =   \sum_{d  \in Y } I_{u,d}   (1-  Q_{u,d})  \end{align*}

Here $ \hat{g}_{u, Y}^b $ represents the \textit{natural} number of  anomalous gaps from the days  not distorted by  too much activity in a station that would give rise to full or empty stations. In contrast, $g^E_{u,Y}$ represents an estimate of anomalous intervals (gaps) in stations owing to excess demand. We assume load imbalance can only occur when there is excess demand, and gaps due to excess demand comprise baseline and baseline gaps. We remove the baseline gaps from gaps owing to excess demand  by subtracting   $ \hat{g}_{u, Y}^b $ from $g^E_{u,Y}$  to refine the estimate of gaps induced by load imbalance. Because load imbalance can only decrease the efficiency of the system by reducing the number of trips,  the demand-correction probability can only be increased and the numerator of \eqref{eq:Pforegone}  is: 
\begin{align}     \label{eq:Eduration1}
\# \{  \text{gaps due to load imbalance in station  } u \text{ in year } Y   \}  
\approx 
\big( \hat{g}^E_{u, Y}  
-   \hat{g}^b_{u, Y}    \big) ^+ 
\end{align} 

The probability of a forgone trip \eqref{eq:Pforegone} can be estimated by   
\begin{align} \label{eq:Probf_re} 
\Prob( \tilde{E}_{u,Y})   &\approx  \frac   {    \big( \hat{g}^E_{u, Y}  
	-   \hat{g}^b_{u, Y}    \big) ^+ 
}     {     \sum_{d \in Y}  (S^*_{u,d}  - 1)   } 
\end{align}

where the denominator, which represents the total number of time-intervals in all days across year $Y$,   can be represented by the sum of trips (daily strengths excluding first and last 10\% of trips) of station $u$ in each day $d$.  We use these probabilities to construct a demand-corrected time-series of graphs $\{\tilde{G}_t\}_{1 \le t \le T}$ and find communities in these networks in addition to the uncorrected graphs. 

\subsection{Trend Testing Details} 

\subsubsection{Testing for Increasing and Decreasing Trends among Node-Sets} \label{app:inc_dec_trend}

For the time trend expressed w.r.t. $t$  given a set $B$, node $v$,
we test for hypotheses for trend about  a symmetric interval $[-U,U]$ close to zero. These hypotheses test for a null hypothesis of zero in equivalence  testing. The null hypotheses are written as follows:
\begin{eqnarray}  
&H_{0,+}:   \beta^+_{ v, B}  \le U   
&H_{1, +}:      \beta^+_{v, B}  > U, 
\\
&H_{0,-}:   \beta^-_{ v, B}   \ge -U  
&H_{1, -}:      \beta^-_{v, B}  < -U. 
\end{eqnarray} 	

We calculate the significance of  $\beta_{v,B}$ using the difference of the estimates as well as the (pre-specified) upper and lower bounds of the trend. In order to filter out the trends that are {negligible}, we perform a t-test for the regression statistic subtracted by the upper or lower bound $U$, divided by the standard error of the estimate, $s(\beta_{v,B})$. Defining such a bound allows us to exclude the  very small but still significant trends and only find clusters that are  increasing or decreasing with considerable magnitude.
\begin{align*}
t_{\text{upper}} (v,B) &=  \frac{ \hat{\beta}^+_{vB}  - U} {s(\beta^+_{vB})} , \quad \quad 
t_{\text{lower}}(v,B) =  - \frac{ \hat{\beta}^-_{vB}  - (-U)} {s(\beta^-_{vB})} 
\end{align*}

{The corresponding p-values of  $  t_{\text{upper}} $ and  $t_{\text{lower}}$, respectively, with significance $\alpha/2$ (for one-sided tests)  {and} with degrees of freedom $n - 2$\lastpass{,} represent the trend of connectivity of node $v$ in relation to  set $B$}. Typical of ordinary least squares, the degrees of freedom are discounted by the slope and intercept terms.

\subsubsection*{Testing for Neutral Trends among Node-Sets}\label{app:neu_trend}
P-values of the similarity of neutral trends to $U$ are obtained by taking the maximum of the p-values associated with the t-statistics $ t_{\text{neutral},a } $ and $ t_{\text{neutral}, b} $, respectively, with significance $\alpha$  and degrees of freedom $n - 2$. 

To determine the t-statistic of a negligible trend, we utilize the approach outlined in \cite{dixon_trend_2008}. To test for whether a trend is negligible, the typical hypothesis test for  a regression coefficient is inverted and split instead into two one-sided tests.

\begin{align*}
H_{0,a}:  \beta_{v,B}\ \ge U , \quad \quad & H_{1,a}:  \beta_{v,B} <  U  ,
\stepcounter{equation}\tag{\theequation}\label{eq:neutral_trendhypothesis_ab}
\\H_{0,b}:   \beta_{v,B} \le  -U  , \quad \quad &  H_{1,b}:    \beta_{v,B} > -U. 
\end{align*}

Dixon et al. (\cite{dixon_trend_2008}) used the following pair of t-statistics  to test for these hypotheses:
\begin{align*}
t_{\text{neutral}, a } &=  \frac{ \hat{\beta}_{uv}  - (-U)}  {  s(\beta_{uv})} ;\quad \quad t_{\text{neutral}, b } =  \frac{ U - \hat{\beta}_{uv}} { s(\beta_{uv})}
\end{align*}
and obtained the corresponding p-values for the probability of the alternative hypothesis by taking the maximum of the p-values associated with the t-statistics $ t_{\text{neutral},a } $ and $ t_{\text{neutral}, b} $, respectively, with significance $\alpha$  and with degrees of freedom $n - 2$.

\subsubsection{Initializing Time Trend of $\xi_{uv, t}$} \label{app:init_TS}

To initialize the iterative search procedure, all individual nodes $u \in 1,...,n$. We calculate $M_0(u)$ for all $B_0(u)=u$ following the procedures from \ref{sec:significant_nodes} at iterative step $k=0$. Within $M_0(u)$,  we calculate each normalized $W_{uv,t } | A_{uv,t}$ by the following equation for all  $v$  that are significantly connected to $u$ across all time $T$:

\begin{align*} 
Z_{t}(u,v) &= \frac{ W_{uv,t } -  \Expec [W_{uv,t} | A_{uv,t}]  }{   \Var (W_{uv,t} | A_{uv,t})  } 
\end{align*} 
where 
\begin{align*} 
\Expec [W_{uv,t} | A_{uv,t}] = \frac{ \frac{s_{u,t} s_{v,t}}  { s_{T,t} }}{\frac{ d_{u,t} d_{v,t}  }{ d_{T,t}} }; \quad 
\Var(W_{uv,t} | A_{uv,t}) =\left( \frac{ \frac{s_{u,t} s_{v,t}}  { s_{T,t} }}{\frac{ d_{u,t} d_{v,t}  }{ d_{T,t}} }\right) ^2 \kappa_t 
\end{align*} 

Next, we find the linear trends of each $Z_{ t} (u,v)$ across time $t=1,...,T$ and take the nodes with trends that are either significantly positive or negative.  We write $\Zb (u,v)  $ as the vectorized time series of $Z_{t}(u,v)$.  The trend is calculated as the coefficient with time $t=1,...,T$ from ordinary least squares (OLS),  between nodes $u$ and $v$. $ \hat{\beta}_{uv} $ is determined to be significantly increasing, decreasing, or stable (neutral) using the method described in the following section \ref{sec:trend_test}, but only using a single node $v$ in place of a set $B.$ If $  \beta_{uv} $ is significant at the $\alpha$ level (in OLS), then denote the nodes $v$ that are significantly connected and  increasing or decreasing with initializing node $u$ as   $v^{**}$. We construct an initializing set $B_1$ with these nodes $\{   u,  \boldmath{v}^{**} \} $ for step $k=1$.

\subsection{Clusters under Alternative Parameterizations}

\subsubsection*{Observed Trips  in Citibike at $\alpha=0.01$}

\begin{figure} [htbp]  	
\centering 
 	\includegraphics 
 	[width=0.85\linewidth, trim={0cm  10cm  0cm 0cm, clip} ]
 	{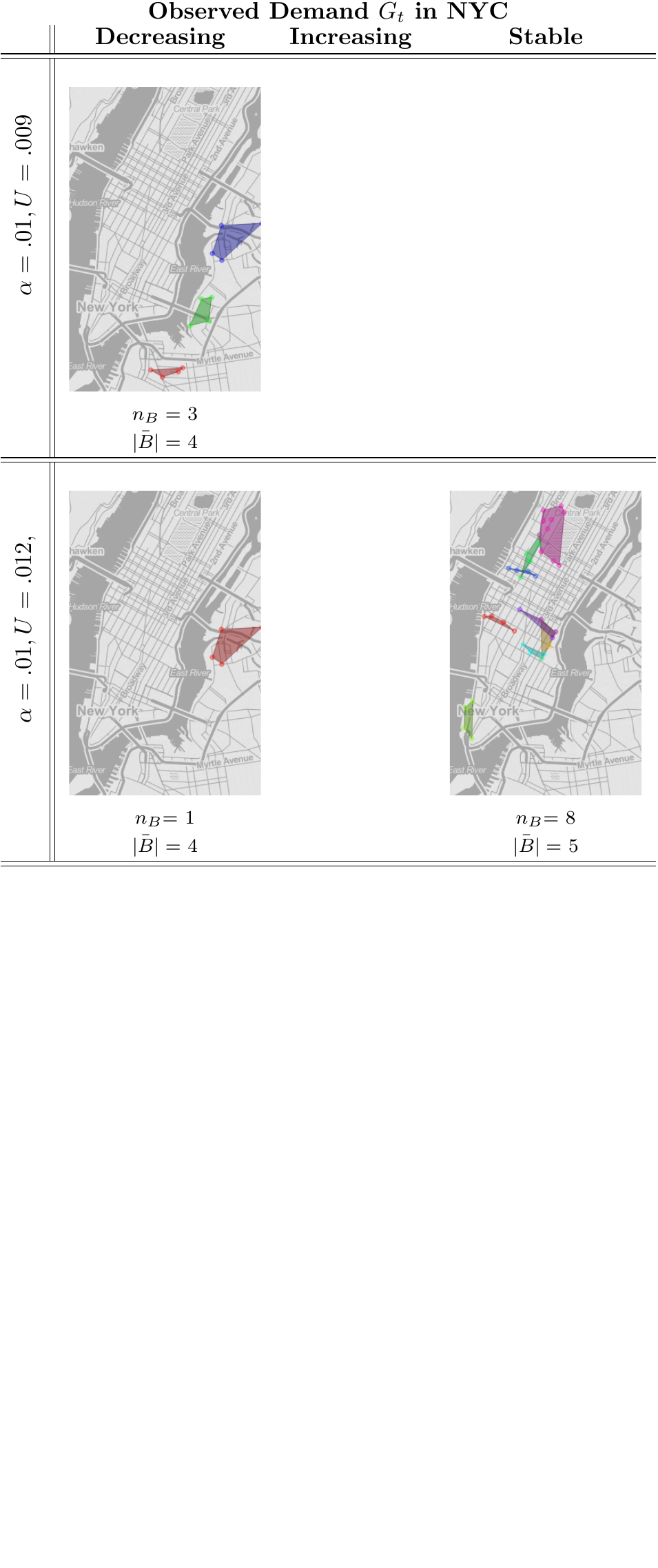}
 	 	
 \caption{Intertemporal Communities of increasing, decreasing, and neutral trends amongst stations in years 2016-2018  in New York City under varying significance levels  and bounding parameters $U$ in the demand-corrected networks $\tilde{G}_t$. $n_B$ represents the number of found communities and $\bar{|B|}$ represents the mean size of communities rounded to the nearest integer.} 
	
	\caption{Intertemporal Communities of increasing, decreasing, and neutral trends amongst stations in years 2016-2018 under varying significance levels  and bounding parameters $U$ in the uncorrected networks $\tilde{G}_t$. $n_B$ represents the number of found communities and $\bar{|B|}$ represents the mean size of communities.   }
	\label{fig:Obs_NYC_alt}
\end{figure}

Several clusters are found in New York city in networks of observed demand $G_t$ (fig. \ref{fig:Obs_NYC_alt}) when $\alpha$ is set to 0.01 as the smallest significance threshold that would allow discovery of clusters in all trend-types . When $U$ is set at 0.009, only decreasing clusters are found. The three decreasing clusters are all located on Long Island in Brooklyn.  When $U$ is set at 0.012, two decreasing clusters disappear but many stable clusters are found. The stable clusters are all located in Manhattan spanning several parts of the island. The remaining decreasing cluster is located in Williamsburg.

\ch{\subsubsection*{Demand-Corrected Networks in New York City}}

\chsb{We apply the intertemporal community detection \jf{algorithm to} the demand-corrected (DC) time-series of networks $ \{ \tilde{G}_t \}_{1\le t \le T}$ with weights $\tilde{W}_{uv,t}$ in the Citibike system.  	
	The obtained communities retain similar geographical characteristics  as communities in uncorrected graphs, but with some differences.
}

When $U$ is set at 0.009,  there are a few decreasing and increasing clusters in the demand-corrected networks but no stable clusters were found. In addition to the clusters found in the non-corrected networks, two clusters in the Upper East Side of Manhattan are found to be decreasing.
The increasing clusters are both located in Central Brooklyn around the {Clinton Hill} neighborhood.  
When  $U$ is increased to 0.012, the increasing clusters vanish and only two decreasing clusters remain. 

\begin{figure}[htbp]
	\centering
	\includegraphics
 	[width=0.85\linewidth, trim={0cm  10cm  0cm 0cm, clip} ]
	{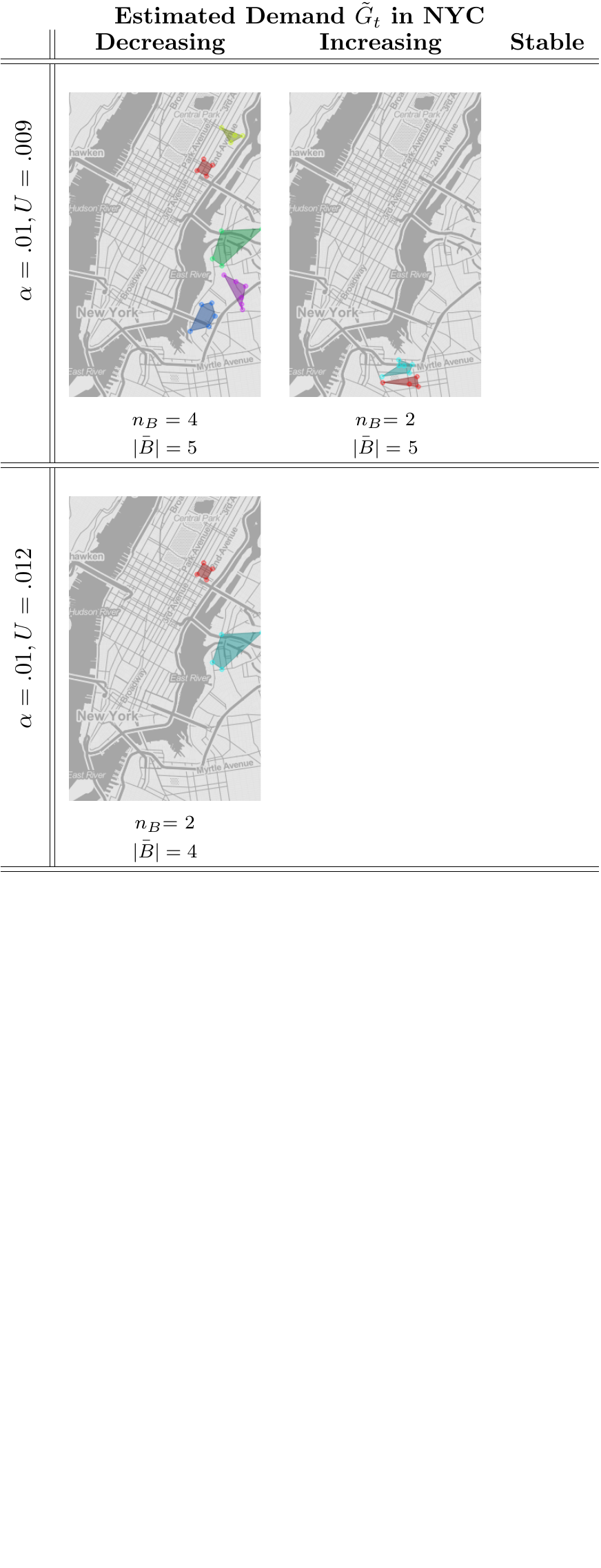}
    \caption{Intertemporal Communities of increasing, decreasing, and neutral trends amongst stations in years 2016-2018  in New York City under varying significance levels  and bounding parameters $U$ in the demand-corrected networks $\tilde{G}_t$. $n_B$ represents the number of found communities and $\bar{|B|}$ represents the mean size of communities rounded to the nearest integer.} 
	\label{fig:aux6appcitibikealt2}
\end{figure}

\subsection*{Acknowledgements}
The authors thank the two referees for an in depth reading of the entire manuscript and detailed comments that lead to a significant improvement of the original submission. 
The authors also thank Eric Hanss for providing helpful information about bikeshare systems, Hannah Loftus for helpful comments on Chicago geography, and Professor Eliza Rose for providing helpful comments on New York geography.

  \subsection*{Availability of Data and Material}
  All the trip data are available on the Divvy website \cite{divvy_data} 
  ({https://www.divvybikes.com/system-data}) and Citibike \cite{citibike}. Reloading data are available from OpenBUS \cite{openbus}.   The code to implement the methods described in this manuscript is available upon request.

\end{document}